\documentclass[12pt,preprint]{aastex}
\usepackage{epsf,emapjbib_only,multicol}
\newcommand{\uv}{\mbox{$u$-$v$}}
\newcommand{\Ho}{\mbox{$H_\circ$}}

\newcommand{\omegam}{\mbox{$\Omega_M$}}
\newcommand{\omegal}{\mbox{$\Omega_\Lambda$}}

\newcommand{\ksM}{\mbox{km s$^{-1}$ Mpc$^{-1}$}}

\newcommand{\Msun}{\mbox{M$_\odot$}}

\newcommand{\muJ} {\mbox{{$\mu\mbox{Jy}$} beam$^{-1}$}}

\newcommand{\asdot}{\mbox{$.^{\prime\prime}$}}

\def\cl1226{Cl\,J1226.9+3332}
\def\clj0152{Cl\,J0152.7$-$1357}
\def\ms1054{MS\,1054.4$-$0321}

\begin{document}

\title{Sunyaev-Zel'dovich Effect Imaging of \\
Massive Clusters of Galaxies at Redshift $z>0.8$}

\author{
Marshall~Joy\altaffilmark{1}, 
Samuel~LaRoque\altaffilmark{2},
Laura~Grego\altaffilmark{3},
John~E.~Carlstrom\altaffilmark{2}, 
Kyle~Dawson\altaffilmark{4},
Harald~Ebeling\altaffilmark{5}, 
William~L.~Holzapfel\altaffilmark{4}, 
Daisuke~Nagai\altaffilmark{2}, \&
Erik~D.~Reese\altaffilmark{2} }
\altaffiltext{1}{Dept. of Space Science, SD50, NASA Marshall Space
Flight Center, Huntsville, AL 35812}
\altaffiltext{2}{Department of Astronomy and Astrophysics, University
of Chicago, Chicago, IL 60637}
\altaffiltext{3}{Harvard-Smithsonian Center for Astrophysics, 60
Garden Street, Cambridge, MA 02138}
\altaffiltext{4}{Physics Department, University of California,
Berkeley, CA 94720}
\altaffiltext{5}{Institute for Astronomy, 2680 Woodlawn Dr., Honolulu,
HI 96822}


\begin{abstract}
We present Sunyaev-Zel'dovich Effect (SZE) imaging observations of
three distant ($z > 0.8$) and highly X-ray luminous clusters of
galaxies, \cl1226, \clj0152 and \ms1054. Two of the clusters, \cl1226
and \clj0152, were recently discovered in deep ROSAT x-ray
images. Their high X-ray luminosity suggests that they are massive
systems which, if confirmed, would provide strong constraints on the
cosmological parameters of structure formation models.  Our
Sunyaev-Zel'dovich Effect data provide confirmation that they are
massive clusters similar to the well studied cluster \ms1054. Assuming
the clusters have the same gas mass fraction as that derived from SZE
measurements of eighteen known massive clusters, we are able to infer
their mass and electron temperature from the SZE data. The derived
electron temperatures are $9.8^{+4.7}_{-1.9}$, $8.7^{+4.1}_{-1.8}$,
and $10.4^{+5.0}_{-2.0}$ keV, respectively, and we infer total masses
of $\sim2 \times 10^{14} h^{-1}_{\rm100}$ \Msun\ within a radius of
65\arcsec\ (340 $h_{\rm100}^{-1}$ kpc) for all three clusters.  For
\clj0152 and \ms1054 we find good agreement between our SZE derived
temperatures and those inferred from X-ray spectroscopy. No X-ray derived
temperatures are available for \cl1226, and thus the SZE data provide the
first confirmation that it is indeed a massive system.  The demonstrated
ability to determine cluster temperatures and masses from SZE observations
without access to X-ray data illustrates the power of using deep SZE surveys
to probe the distant universe.

\end{abstract}


\section{Introduction}
\label{sec:intro}
The existence of galaxy clusters at high redshift can place powerful
constraints on the physical and cosmological parameters of structure
formation models (\cite{bahcall1992, luppino1995, oukbir1997,
donahue1998, eke1998, haiman2000}). The greatest leverage is provided by the
most massive and distant clusters (e.g., \cite{viana1996}).  Two distant and
highly x-ray luminous clusters were recently discovered in deep ROSAT x-ray
images: \cl1226, a cluster at redshift $z=0.89$, was discovered in the WARPS
survey \citep{ebeling2001_cl1226+33,Scharf1997}. \clj0152, at redshift
$z=0.83$, was detected in the RDCS, SHARC, and WARPS surveys
\citep{dellaceca2000, romer2000, ebeling2000_cl0152-13}.  Based on their x-ray
luminosities ($L_{\rm X}[0.5-2 \rm{keV}]
\ga 2 \times 10^{44} h_{\rm100}^{-2}$ erg~s$^{-1}$), these clusters are
thought to be highly massive which, if confirmed, will provide significant
constraints on cosmological models \citep{bahcall1998}.

In this paper, we present interferometric imaging of the
Sunyaev-Zeldovich Effect (SZE) in these clusters, which
provides a measure of the gas pressure integrated along the line of
sight \citep{sunyaev1972, birkinshaw1999}. The change in the observed
brightness temperature of the Cosmic Microwave Background (CMB) 
radiation that results from passage
through the thermally ionized gas permeating a galaxy cluster is given
by 
$${ \Delta T _{thermal}\over T_{CMB}} = f(\nu){k_B \sigma_T \over m_e
c^2}
\int n_e T_e \space dl$$ 
where $T_{CMB}$ is the microwave background temperature (2.7 K);
$\sigma_T$ is the Thomson scattering cross section; and $m_e$, $n_e$,
and $T_e$ are the electron mass, density, and  temperature.  The
frequency dependence of the SZE is represented by $f(\nu)$; in the
Rayleigh-Jeans limit, $f(\nu)=-2$.  We use the SZE to determine the
mass of \cl1226 and \clj0152 using the methods developed by Grego
et al. 2001; in addition, we present interferometric SZE data on
\ms1054, a cluster of known temperature and mass at $z=0.83$
\citep{gioia1994,donahue1998,hoekstra2000}, 
which provides a standard against which the WARPS clusters can be
compared.

The SZE observations and data analysis are described in section II, and
the conclusions drawn from these data will be found in section III.
Throughout this {\it Letter}  we parameterize the Hubble constant in terms
of $h_{\rm100}$,  where $\Ho \equiv 100h_{\rm100}$ \ksM, and we use the
cosmological parameters \omegam=0.3 and \omegal=0.7 unless otherwise stated. 
Uncertainties are reported at the $68\%$ confidence level.

\section{Interferometric Imaging of the Sunyaev-Zel'dovich Effect}
\subsection{Observations}
To image the Sunyaev-Zeldovich Effect in these distant clusters, we
outfitted the Owens Valley Radio Observatory (OVRO) and
Berkeley-Illinois-Maryland-Association (BIMA) millimeter
interferometers with sensitive centimeter-wave receivers optimized for
SZE measurements \citep{carlstrom1996}.  Our receivers use
cryogenically cooled 26-36 GHz high electron mobility transistor
(HEMT) amplifiers \citep{pospieszalski1995}, with characteristic
receiver temperatures of $T_{rx} \sim$11-20 K at the 28.5 GHz frequency used
for these observations. The cluster pointing centers and on-source
integration times are given in Table 1.

The interferometric measurements of  \cl1226, and \clj0152
were made at the BIMA interferometer in 1998 and 2000 with
nine 6.1 meter antennas in a closely packed configuration to maximize
sensitivity to the SZE, with a $6.6\arcmin$ FWHM primary beam and
baselines ranging from 0.6 to 14.3 k$\lambda$ (~6-140 m). Typical system
temperatures, scaled to above the atmosphere, are $\sim$40-45 K in an
800 MHz band centered at 28.5 GHz. Observations of a bright phase
calibrator were interleaved with cluster measurements every 25
minutes, and Mars was used for amplitude calibration
\citep{rudy1987,grego2000}. The MIRIAD software package
\citep{sault1995} was used to calibrate and edit the visibility data
and to output the reduced data in UVFITS format for subsequent
analysis.  

The interferometric measurements of \ms1054 were made at the OVRO
millimeter array in June 1996 with six 10.4 meter antennas in a
closely packed configuration, with a $4.2\arcmin$ FWHM primary beam
and baselines ranging from 1.0 to 12.0 k$\lambda$ (~10-120 m). Typical
system temperatures, scaled to above the atmosphere, are $\sim$45 K in two 1
GHz channels centered at 28.5 and 30.0 GHz (2 GHz total bandwidth).
Observations of a bright phase calibrator were interleaved with
cluster measurements every 24 minutes.  The MMA software
package \citep{scoville1993} was used to calibrate and edit the
visibility data and to output the reduced data in UVFITS format for
subsequent analysis.

We flagged data from baselines when one of the telescopes
was shadowed by another telescope in the array, cluster data that were
not bracketed in time by phase calibrator data (mainly at the
beginning or end of a track), data for which the phase calibrator
indicated poor atmospheric coherence, and, rarely, data with spurious
correlations.

\subsection{Data Analysis}
\label{sec:analysis}
In order to properly model the cluster, we must account for any point
sources in the field.  To identify these point sources, we used
DIFMAP \citep{pearson1994} to produce a high resolution image, using
only data from baselines longer than 20 meters.  The resulting
synthesized beam sizes, RMS noise levels, and point source detections
are given in Table 2.

We perform a quantitative analysis of the observed SZE profiles by
fitting isothermal $\beta$
models \citep{cavaliere1976,cavaliere1981} and point source profiles
to the interferometric data directly in the Fourier ({\it
u,v}) plane, where the noise characteristics and the spatial
filtering of the interferometer are well understood.  The spherical
isothermal $\beta$ model density is described by 
\begin{equation} n_e(r) = n_{e\circ} \left ( 1 + {r^2 \over r_c^2} \right)^{-3 \beta/2},
\label{eq:density}\end{equation} where the core radius $r_c$ and 
$\beta$ are shape parameters, and 
$n_{e\circ}$ is the central electron number density.  With this model,
the SZ Effect temperature decrement is
\begin{equation} {\Delta T}\left ( \theta \right ) = \Delta T_0 \left( 1 +  {\theta^2 \over
\theta_c^2} \right)^{{1\over 2} -{3\beta \over 2}},\label{eq:DT/T}
\end{equation} where $\theta = r/D_A $, $\theta_c = r_c / D_A$, $D_A$ is the angular diameter
distance,  and 
$\Delta T_0$ is the temperature decrement at zero projected radius.

We determine the best-fit point
source positions and fluxes, as well as the cluster centroid,
using a simplex algorithm which minimizes
the chi-squared statistic \citep{reese2000}.  We fix the
cluster centroid and the point source positions and fluxes
at their best-fit values, and calculate the chi-squared
statistic over a large range of $\theta_c$,
$\beta$, and $\Delta T_0$ values.  For a given electron temperature,
$T_e$, the $\beta$ model then yields the gas density profile 
$n_e(\theta$)
at each ($\theta_c$, $\beta$, $\Delta T_0$) point. The gas mass and the
total cluster mass can be calculated directly from $n_e(\theta$),
assuming that the intracluster medium (ICM) is in hydrostatic equilibrium
with the cluster potential.  Following the
methods outlined in Grego et al.\ 2001, we 
calculate the gas mass, total mass, and gas mass fraction 
over the ($\theta_c$,$\beta$, $\Delta T_0$) grid, and report the total mass
values for which the $\chi^2$ statistic is within the 68\% confidence interval
($\Delta\chi^2=1$).  We evaluate these quantities at an angular radius of
65\arcsec, where our observational techniques best constrain the cluster gas
mass fraction \citep{grego2001}.  For clusters at
$z\sim0.8$, this angular radius corresponds to a physical radius of 340
$h_{\rm100}^{-1}$ kpc for an \omegam=0.3, \omegal=0.7 cosmology.

We can estimate the electron temperature directly from the SZE data  by
finding the range of $T_e$ values that yield a cluster gas mass fraction,
$f_g$, consistent with the mean value measured by \cite{grego2001} for a
sample of eighteen clusters. To determine the gas mass fraction at $r_{500}$
for \cl1226, \clj0152, and \ms1054 , we scale the gas mass fractions measured
at 65\arcsec\ to $r_{500}$ using relations derived from numerical simulations
\citep{evrard1996, evrard1997}, as discussed in \cite{grego2001}. This
calculation is repeated for a number of different temperatures ranging from
4-18 keV, and we report the $T_e^{SZ}$ values that are consistent with a mean
value of $f_g(r_{500})=0.081$ \citep{grego2001} within the sample standard
deviation of 0.04. 
To determine how sensitive the derived mass and
temperature are to the adopted cosmological model, we repeated the
calculations above for \omegam=0.3 and \omegal=0.0.  The SZE derived mass and
temperature are reduced by $\sim$3\%; the overall change is small
because the effects of decreased distance are offset by a compensating
change in the $f_g$ scaling relation.

\section{Results and Conclusions}

Synthesized images of the SZ Effect toward \cl1226, \clj0152, and
\ms1054\ are shown in Fig. 1.  The SZE decrement is detected with high
significance in all of these distant clusters, and the locations of the SZE
and X-ray centroids are consistent (Tables 1 and 3). Using the SZE data and
the analysis techniques described in Section ~\ref{sec:analysis}, we
determine the temperature and mass of each cluster (Table 3).   X-ray
temperature measurements for \clj0152 \citep{dellaceca2000} and \ms1054
\citep{donahue1998} are also shown in Table 3, and we find that these X-ray
temperature measurements are consistent with the values inferred from the SZE
data within the stated uncertainties.

From the SZE data, we infer a total mass of $\gtrsim2
\times 10^{14} h^{-1}_{\rm100}$ \Msun\ within a radius of
65\arcsec\space (340 $h_{\rm100}^{-1}$ kpc) for each of the clusters shown in
Table 3.  These mass calculations can be checked against values derived from
gravitational lensing, X-ray, and optical observations of \ms1054
\citep{donahue1998,vanDokkum2000,hoekstra2000}.  \cite{hoekstra2000}
infer a total mass of $5.4\pm0.6 \times 10^{14} h^{-1}_{\rm 100}$ \Msun
within an aperture of radius 94\arcsec\space.  To compare the SZE and
gravitational lensing results, we calculate the gas
mass within a 94$''$ radius aperture, for an electron temperature of 10.4 keV:
$M_{gas}^{SZ}(<94\arcsec)=3.7 \pm 0.6 \times 10^{13} h^{-2}_{\rm100}$ \Msun,
where the uncertainties reflect the statistical 68\% confidence interval in
modelling the SZE data.  The total mass is estimated by scaling this value by
the mean gas mass fraction (Section~\ref{sec:analysis}), from which we obtain
an SZE estimate of the total mass within a 94\arcsec radius:
$M_{total}^{SZ}(<94\arcsec)=4.6\pm 0.8 \times 10^{14} h^{-1}_{\rm100}$
\Msun.  We find that the total mass estimated from the SZE data is consistent
with the lensing measurements of \cite{hoekstra2000}.

Based on the SZE data we conclude that the newly-discovered clusters
\cl1226 and \clj0152 are highly massive, with a total mass of
$M_{total}^{SZ}\gtrsim2 \times 10^{14} h^{-1}_{\rm 100}$ \Msun\ within a radius
of 65\arcsec\ (340 $h_{\rm100}^{-1}$ kpc).  These values are comparable to the
mass inferred from SZE imaging of the $z=0.83$ cluster \ms1054, which has been
confirmed by X-ray, optical, and gravitational lensing studies. These results
demonstrate the ability to determine cluster temperatures and masses from SZE
data without access to  X-ray data, and illustrate the power of using deep SZE
surveys to probe the distant universe. More precise measures of the temperature
and mass of \cl1226 and \clj0152 will be possible with deep X-ray imaging and
spectroscopy, which will be obtained within the coming year by the Chandra and
XMM X-ray observatories; with these data in hand, the SZE measurements can be
used to measure the distance to each cluster \citep{reese2000} and to further
constrain their density, mass, and gas mass fraction \citep{grego2001}.
Additional, independent constraints on the mass distribution in \cl1226 will
be obtained from a weak-lensing analysis of wide-field imaging data taken
with HST and the Subaru 8.3m telescope.

\acknowledgments 
We dedicate this paper to the memory of our friend and colleague Mark Warnock,
who freely gave of his expertise and time and made great contributions to the
interferometric SZE imaging experiment.  We also thank Cheryl Alexander, Rick
Forster, Steve Padin, Dick Plambeck, Steve Scott, David Woody, and the staff of
the BIMA and OVRO observatories for their outstanding support, and Laurence
Jones, Eric Perlman, and Caleb Scharf for providing data on \cl1226 prior to
publication.   ER gratefully acknowledges support from NASA GSRP Fellowship
NGT5-50173. This work is supported by NASA LTSA grants NAG5-7986 (JC,MJ,WH) and
NAG5-8253 (HE).  Radio astronomy at the BIMA millimeter array is supported by
NSF grant AST 96-13998.  The OVRO millimeter array is supported by NSF grant
AST96-13717.  The funds for the additional hardware for the  SZE experiment were
from a NASA CDDF grant, a NSF-YI Award, and the David and Lucile Packard
Foundation.


\bibliographystyle{apj}

\bibliography{mj}

\begin{deluxetable}{lcrrc}

\tablecolumns{5}

\tablecaption{Cluster Observation Log}
\tablehead{
\colhead{Cluster}& \colhead{}& \multicolumn{2}{c}{\underline{Pointing
Center (J2000)}}&
\colhead{On-source Integ.} \\
\colhead{Name} & \colhead{redshift} & \colhead{$\alpha$}
& \colhead{$\delta$} & \colhead{time (hours)} \\
}
\startdata
\cl1226 & 0.89 & 12$^h$26$^m$58$^s$.0 &
33$^{\circ}$32$^{\prime}$45$^{\prime\prime}\tablenotemark{a}$ & 41.6 \\

\clj0152 & 0.83 & 01$^h$52$^m$43$^s$.0	&
 --13$^{\circ}$57$^{\prime}$29$^{\prime\prime}\tablenotemark{b}$ & 27.8 \\

\ms1054 & 0.83 & 10$^h$56$^m$59$^s$.5 &
 --03$^{\circ}$37$^{\prime}$28$^{\prime\prime}\tablenotemark{c}$ & 43.0 \\


\enddata
\tablenotetext{a}{Pointing center coincident with X-ray cluster
center from \cite{ebeling2001_cl1226+33}.}
\tablenotetext{b}{Pointing center coincident with X-ray cluster
center from \cite{ebeling2000_cl0152-13}.}
\tablenotetext{c}{Pointing center coincident with optical
cluster center from \cite{gioia1994}.  The ROSAT/HRI X-ray center is at
10$^h$56$^m$58$^s$.6, --03$^{\circ}$37$^{\prime}$36$^{\prime\prime}$
\citep{neumann2000}.}
\end{deluxetable}

\begin{deluxetable}{lccrrr}

\tablecolumns{6}
\tablecaption{Radio Point Sources}
\tablehead{
\colhead{Cluster} & \colhead{Synth. Beam} & 
\colhead{$\Delta\alpha$} & \colhead{$\Delta\delta$} &
\colhead{Observed} & \colhead{RMS noise} \\

\colhead{Name} & \colhead{($r_{uv}>2 k\lambda$)} &
\colhead{(arcsec)} & \colhead{(arcsec)} &
\colhead{Flux (mJy)\tablenotemark{a}} &
\colhead{(mJy/beam)} \\ }
\startdata
\cl1226 & 14.1\arcsec$\times$16.1\arcsec  & 260.0 &
$-$39.3 & 1.73 & 0.152\\

\clj0152 & 13.7\arcsec$\times$23.1\arcsec & -- & -- &
-- & 0.221\\

\ms1054 & 17.4\arcsec$\times$23.3\arcsec & 0.1 &
1.4 & 0.98 & 0.092\\
& & $-$161.0 & 1.9 & 0.63 & " \\
& & $-$25.5 & $-$86.9 & 0.35 & " \\

\enddata
\tablenotetext{a}{Uncorrected for primary beam attenuation.}
\end{deluxetable}

\begin{center}
\begin{deluxetable}{lrrcccc}

\tablecolumns{5}
\tablecaption{Cluster Properties derived from SZE
measurements}
\tablehead{
\colhead{Cluster} & \multicolumn{2}{c}{\underline{Cluster Position\tablenotemark{a}}} &
\colhead{$T_e^{Xray}$} & \colhead{$T_e^{SZ}$} & \colhead{$M_{total}[<$65\arcsec]}  \\

\colhead{Name} & \colhead{$\Delta\alpha$} & \colhead{$\Delta\delta$} &
\colhead{(keV)} & \colhead{(keV)} &
\colhead{$(10^{14} h^{-1}$\Msun)}\\ }

\startdata
\cl1226 & 0\asdot2 & 12\asdot3 & --  & $9.8^{+4.7}_{-1.9}$ & $2.7\pm0.5$\\

\clj0152 & $-$1\asdot8 & $-$9\asdot2 & $6.5^{+1.7}_{-1.2}$ & $8.7^{+4.1}_{-1.8}$
& $2.1\pm0.7$\\

\ms1054 & $-$7\asdot8 & $-$5\asdot3 & $12.3^{+3.1}_{-2.2}$ & 
$10.4^{+5.0}_{-2.0}$ & $2.3\pm0.3$\\

\enddata
\tablenotetext{a}{Offsets from radio pointing center (Table 1)}
\end{deluxetable}
\end{center}


\clearpage
\begin{figure} 
\epsfysize=7.0in
\epsscale{1.0} 
\plotone{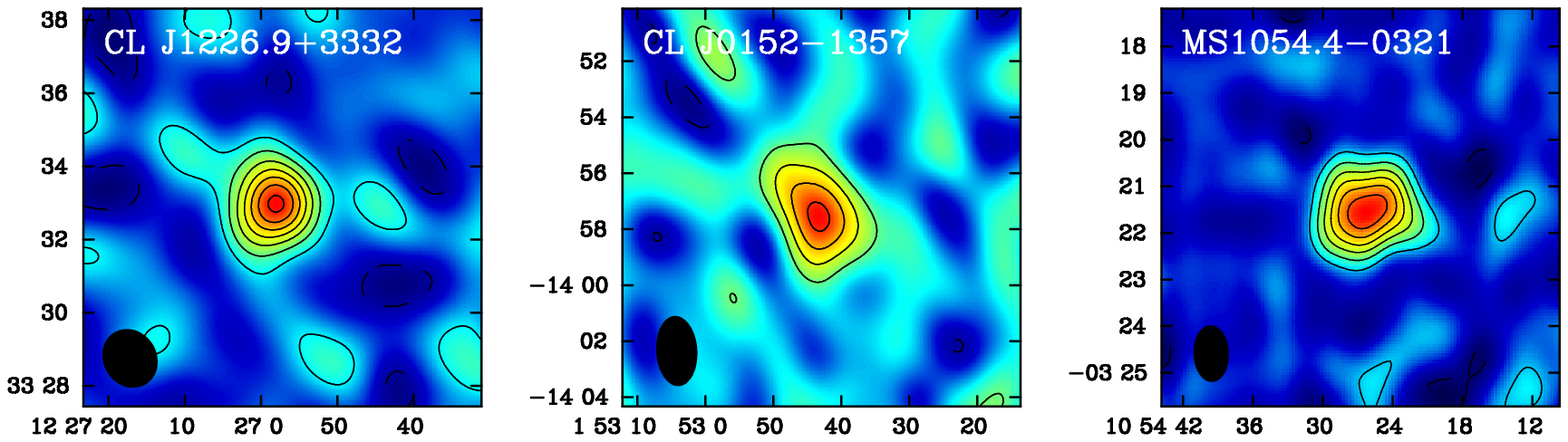}
\caption{Synthesized images of the SZE decrement in \cl1226,
\clj0152, and \ms1054.  
Left panel:  Synthesized image of \cl1226, obtained by applying a Gaussian
taper with a half-power radius of 1 k$\lambda$ to the \uv\space data,
yielding a resolution of $99.3\arcsec \times 87.4\arcsec$ at position angle
(p.a.) $32^{\circ}$.  Contours are multiples of 290 \muJ\space ($1.5\sigma$),
and the rms is 190 \muJ. 
Center panel:  Synthesized image of \clj0152,
obtained by applying a Gaussian taper with a half-power radius of 1
k$\lambda$ to the \uv\space data, yielding a resolution of $151\arcsec \times
87\arcsec.9$ at p.a. $4^{\circ}$.  Contours are multiples of 480 \muJ\space
($1.5\sigma$), and the rms is 320 \muJ.   
Right panel:  Synthesized image of \ms1054, obtained by applying a Gaussian
taper with a half-power radius of 2 k$\lambda$ to the \uv\space data,
yielding a resolution of $73\arcsec.1 \times 45\arcsec.5$ at p.a.
$2^{\circ}$.  Contours are multiples of 120 \muJ\space ($1.5\sigma$), and the
rms is 80 \muJ.} 
\end{figure}

\end{document}